\documentclass[%
reprint,
amsmath,amssymb,
pra,
]{revtex4-2}

\usepackage{graphicx}
\usepackage{dcolumn}
\usepackage{bm}
\usepackage{orcidlink}
\usepackage{braket}
\usepackage{enumitem}
\usepackage{xr} %
\usepackage{color}
\usepackage[version=3]{mhchem}

\begin{document}
	
	\preprint{APS/123-QED}
	
	\title{GUGA-based MRCI approach with Core-Valence Separation Approximation (CVS) for the calculation of the Core-Excited States of molecules}
	
	\author{Qi Song~\orcidlink{0000-0003-4851-3940}}
	\email{songq@nwu.edu.cn}%
	\author{ Baoyuan Liu}
	\author{ Junfeng Wu}
	\author{ Wenli Zou~\orcidlink{0000-0002-2308-2294}}
	\author{ Yubin Wang}
	\author{ Bingbing Suo~\orcidlink{0000-0002-2308-2294}}
	\email{bsuo@nwu.edu.cn}
	\affiliation{1.Institute of Modern Physics, Northwest University, Xi’an, Shaanxi 710069, China}
	\author{ Yibo Lei~\orcidlink{0000-0002-0747-2428}}
	\email{leiyb@nwu.edu.cn}
	\affiliation{2.College of Chemistry and Materials Science, Northwest University, Xi’an, Shaanxi 710069, People’s Republic of China}

	\date{\today}
	
	\begin{abstract}
		We develop and demonstrate how to use the GUGA-based MRCISD with Core-Valence Separation approximation (CVS) to compute the core-excited states. Firstly, perform a normal SCF or valence MCSCF calculation to optimize the molecular orbitals. Secondly, rotate the optimized target core orbitals and append to the active space, form an extended CVS active space, and perform a CVS-MCSCF calculation for core-excited states. Finally, construct the CVS-MRCI expansion space, and perform a CVS-MRCI calculation to optimize the CI coefficients based on the variational method. The CVS approximation with GUGA-based methods can be implemented by flexible truncation of the Distinct Row Table (DRT). Eliminating the valence-excited configurations from the CVS-MRCI expansion space can prevent variational collapse in the Davidson iteration diagonalization. The accuracy of the CVS-MRCI scheme was investigated for excitation energies and compared with that of the CVS-MCSCF method. The results show that CVS-MRCI is capable of reproducing well-matched vertical core excitation energies that are consistent with experiments, by combining large basis sets and a rational reference space. The calculation results also highlight the fact that the dynamic correlation between electrons makes an undeniable contribution in core-excited states.
	\end{abstract}
	
	\maketitle
	
	
	\section{\label{sec:level1}INTRODUCTION}
	Core-excited states, by pumping the inner-shell electrons (usually the \textit{K} shell for the first-row elements, and the \textit{K} and \textit{L} shells for the second-row elements) into unoccupied orbitals, always lie in the high-energy region or even in the X-ray region. Compared to ground and valence-excited states, core-excited states can reveal greater details, such as identifying the origin of specific experimental signatures, and extracting detailed electronic and structural information \cite{Martin-Diaconescu2007,Wang2017}.
	
	Experimental X-ray spectroscopy characterization has now become a scientific discipline thanks to the emergence of second-generation synchrotron radiation facilities \cite{Emma2010,Allaria2012,Ishikawa2012,Zholents1996,Rousse2004,Weigold2000,Li2010,Mcneil2010,Rohringer2012,Pellegrini2016}. The third-generation synchrotron radiation facilities' ongoing construction and the development of fourth-generation facilities (X-ray free-electron lasers, or XFELs) surely provide a more sensitive experimental tool\cite{Khakhulin2020,Milne2014}. It is indisputable that X-ray absorption spectrum (XAS) is now an indispensable tool in many branches of natural research \cite{Young2018}.
	
	Besides the rapid development of experimental facilities, theoretical approaches to study the core-excited states have undergone comparable advances \cite{Rankine2021}. As early as 1965, Bagus had proposed a {$\Delta$}-SCF (Self-Consistent-Field) approach to determine the core-excited state.\cite{Bagus1965} In recent years, a number of wave function approaches were carried out to study the core excitations, such as the second-order complete active space perturbation theory (CASPT2)\cite{Ljubic2014,Su2016}, the equation of motion coupled cluster theory (EOM-CCSD)\cite{Liu2019,Nascimento2017,Peng2015,Coriani2015,Nooijen1995}, and  the algebraic diagrammatic construction (ADC)\cite{Trofimov2000,Wenzel2014,Wenzel2016,Wenzel2014-2,Wenzel2015}. But the majority of computational studies have been carried out using the time-dependent density functional theory (TD-DFT) due to its low computational cost \cite{Pople1992,Besley2010,DeBeerGeorge2008,Akama2010,Lopata2012,Ljubic2016,Stener2003}. However, to achieve agreement with experiments, it is always necessary to add an energy shift to the calculated spectrum.
	
	This work focuses on how to describe the core-excited states accurately with the multi-reference configuration interaction method with single and double excitations (MRCISD or MRCI for short). The main advantage of MRCI is its ability to yield accurate and reliable descriptions of the electronic structures when an appropriately large basis set and a reasonable reference space are emplyed. Due to a bottom-up approach adopted by the traditional eigenvalue solvers\cite{Davidson1975}, all energetically lower-lying excited states should be obtained before core excitations are targeted. Even though the computational cost can be significantly reduced by using appropriate algorithms, such as hole-particle symmetry\cite{Zhen-yi1982,Suo2018} and a flexible internal contraction scheme,\cite{Shamasundar2011,Werner1988,Wang2014,Werner1982,Lee1987,Celani2000,Wang2004,Knowles1992,Siegbahn1983} for a larger configuration interaction space, MRCI calculation remains a time-consuming work \cite{Wang2014}. On the other hand, the core-excited states, which locate in the high-energy X-ray region of the electronic spectrum,introduce various issues that become apparent in the context of core-excited states. It is unrealistic to calculate the core-excited states by MR methods with iterative diagonalization without introducing some approximations to the existing MRCI scheme.
	
	As a reasonably simple yet effective solution to these challenges, the core-valence separation approximation (CVS) scheme, originally proposed by Cederbaum, Domcke, and Schirmer in the context of Core-Ionized in 1987 \cite{Cederbaum1987}, was adopted for the computations of core-excited states. Since core orbitals are strongly spatially localized around the corresponding atoms, the energy and spatial localization differences between core and valence orbitals are so significant that the couplings between core and valence-excited states are very small and can be ignored. The electronic Hamiltonian matrix can then be effectively reduced to taking into account only core-excited states. The core-excited states and the corresponding wave functions can then be derived by diagonalizing this low-dimensional Hamiltonian matrix using an iterative or direct diagonalization scheme. Another noteworthy issue is the relativistic effect. Since inner-shell electrons are attracted by a stronger Coulomb interaction and bounded by the nucleus, they are more susceptible to relativistic effects than valence electrons. Relativistic effects result in a much more significant lowering of the energy of the core orbitals than the valence orbitals.
	
	\section{\label{sec:level1}THEORY AND IMPLEMENTATION}
	
	\subsection{\label{sec:level2}MRCISD and CVS-MRCISD}
	
	The MRCISD wave function is written as
	\begin{eqnarray}
		\left | \varPsi_{MRCI}  \right \rangle  &=&\sum_{r=1}^{N_{\text {ref }}} c_{r}\left|\phi_{r}\right\rangle+\sum_{r=1}^{N_{\text {ref }}} \sum_{i, a} c_{r}^{i a}\left|\phi^{i a}_{r}\right\rangle\nonumber \\
		& &+\sum_{r=1}^{N_{\text {ref }}} \sum_{i>j, a>b} c_{r}^{ij,ab}\left|\phi^{i j, a b}_{r}\right\rangle
	\end{eqnarray}
	where  ${\ket{\phi_{r}},(r=1,N_{\text {ref}})}$ are the reference configurations. $\ket{\phi^{ia}_{r}}$ and $\ket{\phi^{ij,ab}_{r}}$ are the singly and doubly excited configurations respect to $\ket{\phi_{r}}$, respectively. 
	Here, the orbital indices $\textit{i}$, $\textit{j}$ and $\textit{a}$, $\textit{b}$ range over the occupied and unoccupied orbitals in the reference configuration $\ket{\phi_{r}}$, respectively. Since a particular excitation from one reference configuration might be identical to a different excitation from some other reference configurations, it is necessary to eliminate these redundant expansion configurations. $c_{r}$, $c_{r}^{ia}$ and $c_{r}^{ijab}$ are the expansion coefficients, which can be determined by solving the eigenvalue equation,
	
	\begin{eqnarray}
		\textit{HC}=\textit{EC}
	\end{eqnarray}
	Here, the matrix element of the Hamiltonian is written as:
	\begin{eqnarray}
		\left\langle\Phi_{\mu}|H| \Phi_{v}\right\rangle&&=\sum_{p, q}\left(p\left|h_{1}\right| q\right)\left\langle\Phi_{\mu}\left|E_{p q}\right| \Phi_{v}\right\rangle\nonumber \\
		& & +\frac{1}{2} \sum_{p, q, r, s}\left(p q\left|h_{12}\right| r s\right)\left\langle\Phi_{\mu}\left|E_{p q, r s}\right| \Phi_{v}\right\rangle\label{eqn:coupcoeff}
	\end{eqnarray}
	In which $\left(p\left|h_{1}\right| q\right)$ and $\left(p q\left|h_{12}\right| r s\right)$ are one- and two-electron integrals in the Mulliken notation, respectively. The excitation operators $\textit{E}_{pq}$ and $\textit{E}_{pq,rs}$ are expressed as,
	\begin{eqnarray}
		E_{p q}=\sum_{\sigma} a_{p \sigma}^{\dagger} a_{q \sigma}
	\end{eqnarray}
	\begin{eqnarray}
		E_{p q, r s}=E_{p q} E_{r s}-\delta_{q r} E_{p s}
	\end{eqnarray}
	where $ a_{p \sigma}^{\dagger}$ and $a_{q \sigma}$ are the creation and annihilation operators for an electron on orbitals $p$ and $q$ with spin $\sigma$, respectively.
	
\begin{figure*}
		\includegraphics[angle=0,width=1.0\textwidth]{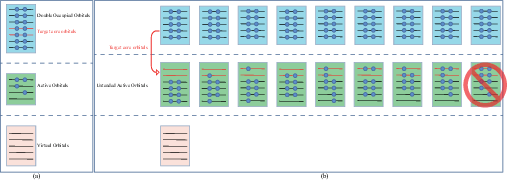}
		\caption{\label{fig:wide}(a) The orbital scheme used to construct the MR wave function. (b) The orbital scheme used for the construction of the CVS-MR wave function. The optimized target core orbitals are rotated and appended to the active space, forming an extended active space. Only a few typical configurations are enumerated here. The red lines in the green background boxes represent the optimized target core orbitals. Electronic configurations with fully occupied core orbitals are eliminated in CVS-MCSCF.}
		\label{Figure1}
	\end{figure*}

	According to the occupation pattern in the reference states, molecular orbitals can be conveniently divided into doubly occupied, active, and virtual orbitals. The doubly occupied orbitals are also called hole orbitals according to the hole-particle symmetry \cite{Zhen-yi1982}. The molecular orbitals are arranged in ascending order from 1 to $\textit{n}$ as virtual, active, and hole orbitals. Here, $\textit{n}$ is the total number of correlated orbitals. The hole and active orbitals constitute the internal space, while the virtual orbitals make up the external space. The orbital scheme for both the construction of the multi-reference (MR) (a) and CVS-multireference (CVS-MR) (b) wave functions is shown in FIG. \ref{Figure1}, and the notations employed are listed in TABLE \ref{tab:Table1}.
	\begin{table}[b]
		\caption{\label{tab:Table1}
			Notations used in this work.}
		\begin{ruledtabular}
			\begin{tabular}{lp{6cm}}
				\textit{Notation}  & \\
				\hline
				\textit{I, J} & Indices of the target core orbitals \\
				\textit{i, j} & Indices of the occupied orbitals in the reference states  \\
				\textit{a, b} & Indices of the unoccupied orbitals in the reference states  \\
				\textit{p, q, r, s} & Indices of the arbitrary orbitals  \\
				\textit{$\overline{X}\underline{Y}$} & A sub-DRT in the active space from the head-node $\overline{X}$ to the tail-node $\underline{Y}$. \\
				\textit{$\Phi_{\mu}$, $\Phi_{\nu}$} & Configuration state functions (CSFs) or Gelfand states  \\
				\textit{$\phi_{r}$} & Reference configuration functions of active orbital space \\
				\textit{$\phi_{0}$} & Reference configuration function of CVS-active orbital space \\
				\textit{$\varphi_{i}$} & The \textit{i}th molecular orbital  \\
				\textit{$\ket{d_1d_2...d_r...d_n}$} & A step vector in GUGA \\
			\end{tabular}
		\end{ruledtabular}
	\end{table}

	A generally used scheme to define the reference configurations is complete active space (CAS), in which the reference configurations with the required spin and spatial symmetry are generated by a given number of electrons distributed in all possible ways in a given set of molecular orbitals. We should start with CVS-CASSCF and use it to generate the reference space for the CVS-MRCISD calculation. In this approach, the molecular orbitals of the ground state are optimized by first running the CASSCF calculation with \textit{n} active electrons distributed in \textit{m} valence active orbitals (denoted as CAS(\textit{n},\textit{m})). The active orbitals are depicted in FIG. \ref{Figure1}(a). Subsequently, \textit{M} optimized targeted core orbitals (TCOs) are rotated into the active space, forming a restricted active space [RAS(\textit{n}+2\textit{M},\textit{m+M})] in which at least one electron should be excited out of the TCOs as illustrated in FIG. \ref{Figure1}(b). Therefore, all electron configurations in the RAS are core-excited. We use RAS rather than CAS because TCOs fully occupied electron configurations have lower energy, which causes the Davidson diagonalization to collapse into the ground or valence excited states. The linear combination of core-excited configurations in RAS forms a core-excited state as
	\begin{eqnarray}
		\ket{\Phi_{mc}^{CVS}}=\sum_{I=1}^M\sum_{a}c_I^a\ket{\phi_{I}^a}+\sum_{I>J}^M\sum_{a>b}c_{IJ}^{ab}\ket{\phi_{IJ}^{ab}}+...\label{Eq8}
	\end{eqnarray}
	with 
	\begin{eqnarray}
		\begin{aligned}
			&\ket{\phi_{I}^a}=E_{Ia}\ket{\phi_0} \\
			&\ket{\phi_{IJ}^{ab}}=E_{Ia,Jb}\ket{\phi_0} \\
			&...
		\end{aligned}
	\end{eqnarray}
	contain all possible excitations from TOCs to the valence active orbitals. Moreover, 
	\begin{eqnarray}
		\ket{\phi_0}=\ket{\varphi_1^2\varphi_2^2...\varphi_I^2\varphi_J^2...\varphi_M^2\phi_{r}} (r=1,{N_{\rm CAS(n,m)}})
	\end{eqnarray}
	are electron configurations generated by CAS(\textit{n},\textit{m}) for the ground state, with which \textit{M} TCOs are double occupied. \textit{I, J} and \textit{a, b} are the indices of TCOs and valence active orbitals in RAS, respectively. \textit{c$_{I}^a$}, \textit{c$_{IJ}^{ab}$} in Eq. \ref{Eq8} are the configuration coefficients. These configuration coefficients for the core-excited states are optimized using a CVS-MCSCF calculation in this RAS space. Targeted core orbitals are frozen and fixed in the extended active orbital space during the CVS-MCSCF calculation to prevent them from rotating back into the doubly occupied orbital space. 
	
	In CVS-MRCISD, the core-excited configurations on the right-hand side of Eq.\ref{Eq8} are selected as reference configurations. It is critical to prevent exciting electrons into the unoccupied TCOs, otherwise, it will cause variational collapse during the CVS-MRCISD calculation. Such restriction is implemented by modifying the Distinct Row Tableau (DRT) in the Graphical Unitary Group Approach (GUGA) as discussed in the next section.
	
	
	\subsection{\label{sec:level2}GUGA and DRT}
	GUGA was proposed by Shavitt,\cite{Shavitt1977,Shavitt1978} which provided not only a compact way to record configuration state functions (CSFs) of CI calculation in a DRT, but also a simple method to calculate the coupling coefficients in Eq. \ref{eqn:coupcoeff}. In GUGA, CSFs are Gelfand states, represented as Paldus tableaus and recorded in DRT.\cite{Paldus1980,Paldus1974,Shepard2020} DRT consists of nodes and arcs,where DRT nodes on the \textit{r}th orbital are distinct rows in all Paldus tableaus on the \textit{r}th row. Three positive integers $(a_r, b_r, c_r)$ are used to label a DRT node, where $a_r, b_r$ and $c_r$ are the numbers of doubly occupied, singly occupied, and virtual orbitals in the first \textit{r} orbitals in a Gelfand state with
	\begin{equation}
		a_{r}=\frac{1}{2}N_r-S_r
	\end{equation}
	\begin{equation}
		b_{r}=2S_r
	\end{equation}
	\begin{equation}
		r=a_r+b_r+c_r
	\end{equation}
	where $N_r$ and $S_r$ are the number of electrons and the corresponding total spin in the first $\textit{r}$ orbitals, respectively.
	An arc links the node $(a_r, b_r, c_r)$ on the \textit{r}th row and $(a_{r-1}, b_{r-1}, c_{r-1})$ on the (\textit{r-1})th row is called a \textit{step} and is denoted by the integer $\textit{d}_r$ with
	\begin{equation}
		d_r=3(a_r-a_{r-1})+(b_r-b_{r-1}).
	\end{equation}
	All allowed steps linked to two DRT nodes are summaries in TABLE \ref{tab:Table2} \cite{Shamasundar2011,Suo2018,Shepard2020} An individual Gelfand state is a complete walk from the DRT's head to the tail as
	\begin{align}
		|\Phi_{\mu}\rangle =& |((d_v)(d_a)(d_h))_{\mu}\rangle \\
		(d_v) =& d_1 d_2 \cdots d_{n_v}      \nonumber        \\
		(d_a) =& d_{n_v+1} d_{n_v+2} \cdots d_{n_v+n_a}    \nonumber \\
		(d_h) =& d_{n_v+n_a+1} d_{n_v+n_a+2} \cdots d_{n}  \nonumber
	\end{align}
	Here, $n_v$, $n_a$ and $n_h$ are the number of virtual, active and doubly occupied orbitals, respectively. $n$ is the number of all correlated orbitals. 
	
	\begin{table}[htpb]
		\caption{\label{tab:Table2}
			Step description. All $\Delta$ quantities are of the form $\Delta X=X_r-X_{r-1}$.}
		\begin{ruledtabular}
			\begin{tabular}{cccccc}
				\textit{d} & $\Delta a$ & $\Delta b$ & $\Delta c$ & $\Delta S$ & $\Delta N$  \\
				\noalign{\smallskip}\hline\noalign{\smallskip}
				0 & 0 & 0 & 1 & 0 & 0\\  
				1 & 0 & 1 & 0 & $\frac{1}{2}$ & 1  \\
				2 & 1 & -1 & 1 & $-\frac{1}{2}$ & 1  \\
				3 & 1 & 0 & 0 & 0 & 2  \\
			\end{tabular}
		\end{ruledtabular}
	\end{table}
	
	Our GUGA implementation fully exploits the hole-particle symmetry, which not only provides a straightforward classification scheme for CSFs but also streamlines the coupling coefficient calculation. Note that there are two boundaries in a complete DRT, one is the boundary between the hole and the active space (the hole-active boundary), and the other is the boundary between the active and the virtual space (the active-virtual boundary). Considering the number of electrons on the virtual orbitals and the spin coupling mode, the nodes of DRT at the active-virtual boundary are classified into  $\underline{V}$(no electron), $\underline{D}$(one electron, doublet), $\underline{T}$(two electrons, triplet), and $\underline{S}$(two electrons, singlet). According to hole-particle symmetry, the nodes of DRT at the hole-active boundary belong to  $\overline{V}$(no hole), $\overline{D}$(one hole, doublet), $\overline{T}$(two holes, triplet), and $\overline{S}$(two holes, singlet), respectively.  In the active space, the fragment of DRT naturally divides into sub-DRTs labeled by $\overline{X}_\sigma \underline{Y}_{\sigma'}$, which is a collection of nodes and arcs with the same head node $\overline{X}_\sigma$ and tail node $\underline{Y}_\sigma'$, respectively. Here, $\sigma$ and $\sigma'$ denote irreducible representations when the molecular point group is used in the calculation. For a given $\overline{X}_\sigma \underline{Y}_{\sigma'}$, the CSFs belonging to this space have definite numbers of holes and particles.\cite{Suo2018}  For instance, Gelfand states passing through the sub-DRT($\overline{V}\underline{V}$) are the reference configurations, in which there are no holes and electrons in the doubly occupied space and the virtual space. Subspace $\overline{D}\underline{S}$ means that there is one hole on the doubly occupied orbital and two electrons coupled as a singlet state on the virtual orbitals. When hole-particle symmetry is used, only sub-DRTs in active spaces need to be generated in calculation. The core-valence separation in MRCISD can be achieved by modifying the sub-DRTs in active space as we will demonstrate later.
	
	The value of a coupling coefficient can be calculated as the product of the segment factors on each orbital levels in a complete loop as, \cite{GUGA}
	\begin{align}\label{eqn:loopie}
		\Gamma^{\mu\nu}_{pq,rs} =& \langle ((d_v)(d_a)(d_h))_\mu|E_{pq,rs}|((d_v)(d_a)(d_h))_\nu\rangle \\
		&= ELS*ALS*HLS  \\
		ELS =& \sum_{J=0,1}\omega_J\prod^{n_v}_{x=M} W(Q_x,d^\prime_xd_x,\Delta b_x,b_x,J) \\
		ALS =& \sum_{J=0,1}\omega_J\prod^{n_v+n_a}_{x=n_v+1} W(Q_x,d^\prime_xd_x,\Delta b_x,b_x,J) \\
		HLS =& \sum_{J=0,1}\omega_J\prod^{M^\prime}_{x=n_v+n_a+1} W(Q_x,d^\prime_xd_x,\Delta b_x,b_x,J) 
	\end{align}
	Here, the coupling coefficient has been written as the product of three factors named $ELS$, $ALS$ and $HLS$, each of which is the product of segment factors of a complete loop in three distinct orbital spaces, respectively. The values of the segment factors $W(Q_x,d^\prime_xd_x,\Delta b_x,b_x, J)$ are determined by the segment type $Q_x$, the step of bra $d_x$ and ket $d^\prime_x$, $\Delta b_x = b_x-b^\prime_{x}$ and $b_x$ value of the ket vector. $J$ is used to account for the spin coupling between two generators $E_{pq}$ and $E_{rs}$. $\omega_J$ may have a value of 0, 1 and -1 depending on whether the two generator lines intersect. In case two lines intersect, $\omega_J=-1$; if not, $\omega_J=1$. $\omega_J=0$ if  If $E_{pq}$ and $E_{rs}$, have no orbital overlap area.
	
	 All segment factors have been tabulated in Ref. \citep{GUGA}. Because only one or two electrons are allowed to excite in the MRCISD calculation, DRT in the hole and virtual spaces have a simple graphical structure. All $ELSs$ and $ALSs$ can be deduced in advance, while $ALSs$  are calculated on the fly in the DRT fragment in the active space.\citep{DirectCI,HoleParticalSymmetry}
	
	\subsection{\label{sec:level2}Realization of Core-Valence Separation (CVS) with DRT}
	
	\begin{figure*}
		\includegraphics[angle=0,width=1.0\textwidth]{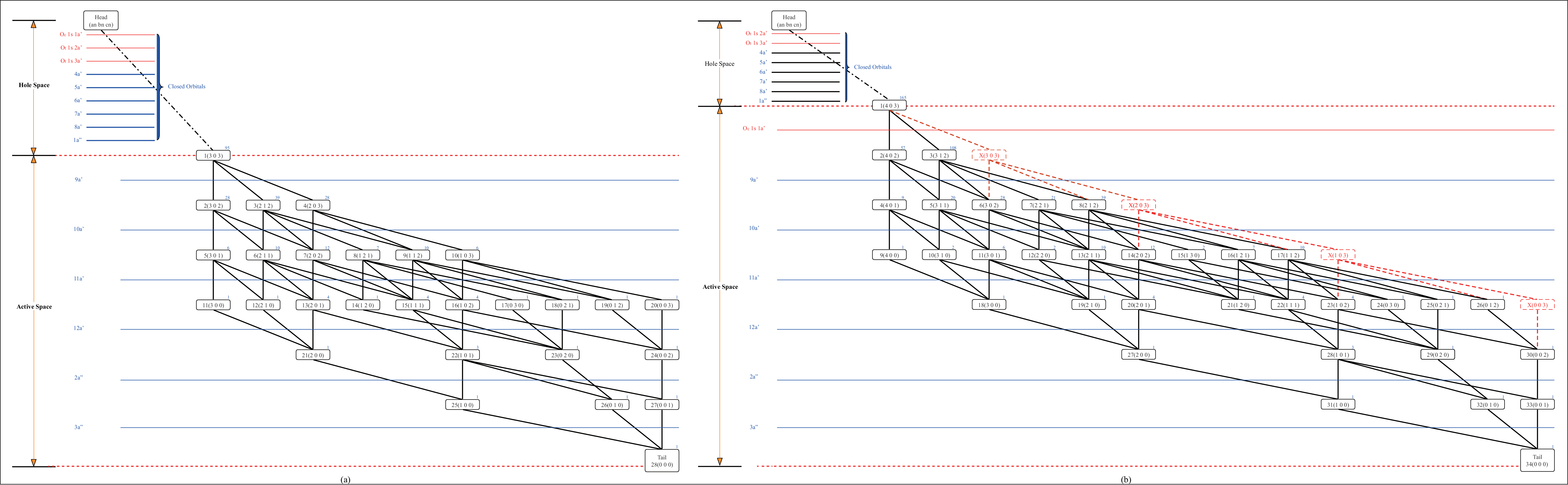}
		\caption{\label{fig:wide}DRT for $O_3$ /CAS (6, 6)/Cs/aug-cc-pVTZ. (a) The graph describing of the reference configuration space (sub-DRT $\overline{V}\underline{V}$) and (b) The graph describing of the CVS-reference configuration space.}
		\label{Figure2}
	\end{figure*}
	
	The key point of the CVS approximation is to separate the core excitation configurations from the lower energy valence excitation configurations. Only the core-excited configurations are taken into account in the CVS-MRCI calculation. As can be seen in TABLE \ref{tab:Table2}, the step vector $\textit{d}_r=3$ indicates that the $\textit{r}$th orbital is doubly occupied. Therefore, Gelfand states with the step "3" on all TCOs in extended active space are valence-excited configurations, which should be eliminated in accordance with the CVS approximation. This work is accomplished by omitting some nodes and arcs in sub-DRTs of the extended active space.
	
	FIG. \ref{Figure2} depicts how we modify the sub-DRT to eliminate the valence-excited configurations and achieve CVS. The reference configuration space (labeled as $\overline{V}\underline{V}$) is chosen as a restricted active space with 8 electrons distributed on 7 orbitals (denoted as RAS(8,7)) for \ce{O3} with $C_{2v}$ symmetry. The first orbital in the extended active orbital space is the $1\textit{s}$ core orbital of the intermediate O atom, which is the TCO. Electronic configurations with $d_1=3$ in the sub-DRT $\overline{V}\underline{V}$ are eliminated, as indicated by dashed lines, which implies that the target core orbital should be empty or singly occupied. The CVS-MRCI expansion space is constructed by applying the single- and double-excitation operators to the CVS-reference configurations, simultaneously eliminating the non-core excitation and redundant configurations.
	
\begin{figure*}
		\includegraphics[angle=0,width=1.0\textwidth]{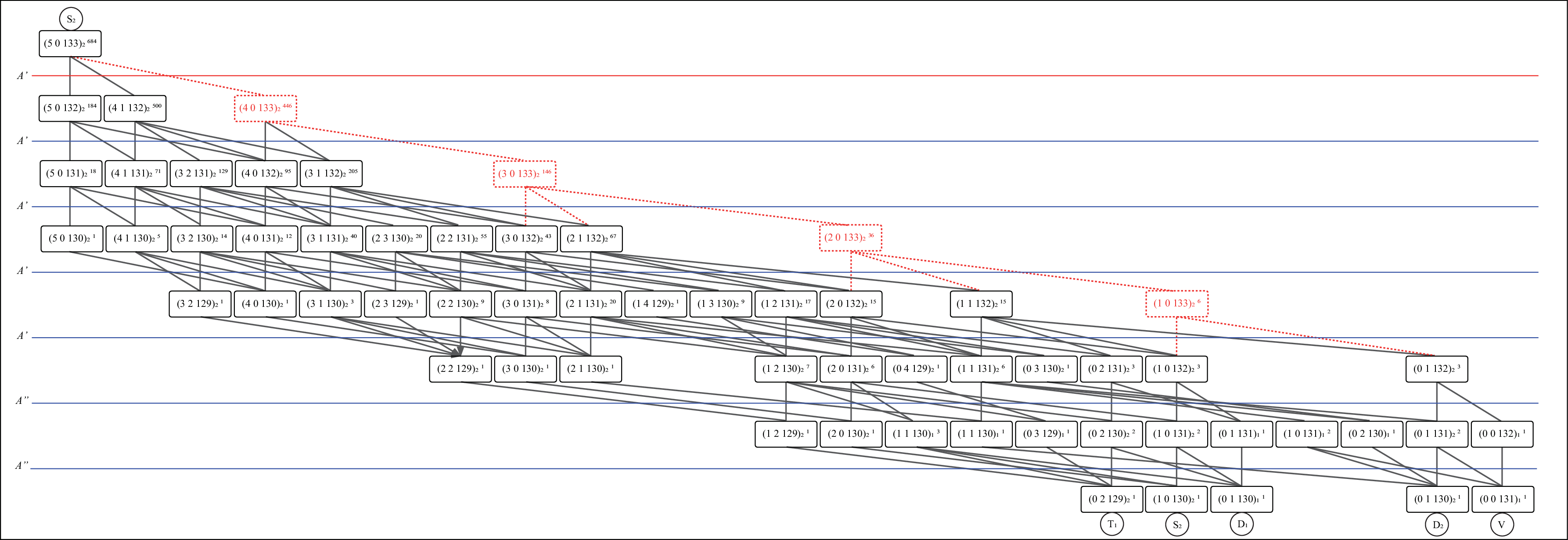}
		\caption{\label{fig:wide}The sub-DRTs $\overline{S}_2\underline{X}$ ($\underline{X}=\underline{V},\underline{D}_1,\underline{D}_2,\underline{T}$, respectively). The nodes (in distinct rows) are indicated by $(a, b, c)_\gamma$, the subscript $\gamma$ is the symmetry index (1-A’,2-A”), and the superscript is the down-step weight of the node. The red dashed lines and the nodes in the red dashed boxes are eliminated in the CVS approximation.}
		\label{Figure3}
	\end{figure*}

	FIG. \ref{Figure3} shows how to construct the CVS excitation space using only sub-DRTs $\overline{S}_2\underline{X}$ ($\underline{X}=\underline{V}, \underline{D}_1, \underline{D}_2, \underline{T}$) as an example, and compares the sub-DRTs with and without the CVS approximation. Then the eigen-energy of the core-excited states can be obtained by solving the eigenequation using either the iterative diagonalization method or the direct diagonalization method.
	
	The CVS approximation with GUGA-based methods can be easily implemented as follows:
	\begin{itemize}[itemsep=0pt,topsep=0pt,parsep=0.1pt]
		\item [a)] \textit{Optimize the MOs (including the target core orbitals in the hole space) for the ground state;}
		\item [b)] \textit{Rotate the optimized target core orbitals and appended into the active space, perform a CVS-MCSCF calculation for core-excited states, and generate CVS reference configurations;}
		\item [c)] \textit{Invoke the single- and double-excitation operators on the CVS-reference configurations and  eliminate the redundant and non-core excitation configurations;}
		\item [d)] \textit{Perform a CVS-MRCI calculation for the core-excited states with variational approach to obtain the eigenvalues and corresponding eigenfunctions of these states.}
	\end{itemize}
	
	\section{\label{sec:level1}COMPUTATIONAL DETAILS}
	
	All CVS-MRCI calculations were carried out for a range of basis sets. Nevertheless, the core-valence correlation described by these basis sets was found to have a negligible effect on the simulated core-excited spectra. Ground state geometry optimizations were performed using multi-reference second-order perturbation theory (MRPT2) with aug-cc-pVTZ basis set, as this combination was shown to provide accurate structures. All calculations were carried out with the BDF program package.\cite{Zhang2020} In particular, both the CASSCF and Xi’an-CI modules in BDF were used to perform the core-excited multi-states calculations.\cite{Wang2014,Lei2012,Lei2021} Orbital localization is necessary for some molecules with high symmetry, as the inner shell orbitals are delocalized over the corresponding atoms. The effect of relativistic effects computed with the relativistic aug-cc-pVTZ-DK basis sets, with the Douglas-Kroll-Hess second-order scalar relativistic correction, is considered.
	
	\section{\label{sec:level1}RESULTS and DISCUSSION}
	The relativistic effect can significantly reduce the energy of core orbitals. The lower the principal quantum number \textit{n} of the core orbital, the more severe the relativistic correction to the orbital energy. In this work, the scalar relativistic corrections are computed at the HF or MCSCF levels with the aug-cc-pVTZ basis set. The results show that the energy reduction is insensitive to the methods adopted. 
	
	Our results show that the average core orbital energy corrections due to relativistic effects are -0.114 eV for the carbon $1\textit{s}$ orbital, -0.224 eV for the nitrogen $1\textit{s}$ orbital, -0.408 eV for the oxygen $1\textit{s}$ orbital and -0.676 eV for the fluorine $1\textit{s}$ orbital. The relativistic correction averages of the $1\textit{s}$ orbitals in the K shell of the Si, P, S, and Cl atoms in the second row are -4.455 eV, -5.971 eV, -7.860 eV, and -10.172 eV, respectively. As expected, this correction becomes more significant as the atomic number rises, although this effect also slightly depends on the molecule in which the atom is located.

	\subsection{\label{sec:level2}Vertical Core-Excited States}
	
\begin{figure}[b]
		\includegraphics{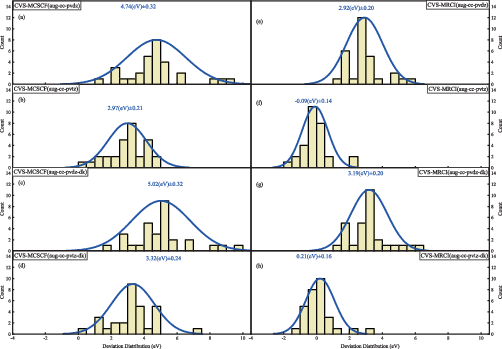}%
		\caption{\label{Figure5} The deviation distribution of the computed excitation energies for the first-row elements.}
	\end{figure}

	The calculated vertical core excitation energies for the first-row elements (C, N, O, and F) in optimized structures of a group of small molecules are listed in TABLE \ref{SI:table1} \cite{SI}. Related experimental values for reference are collected from previous works.\cite{Tronc1979,Besley2009,Asmuruf2008} The deviation between the calculated and experimental values, as well as the deviation distribution, are illustrated in FIG. \ref{Figure4} and FIG. \ref{Figure5}, respectively. From FIG. \ref{Figure5}, one of the most notable features of the results is the poor performance of CVS-MCSCF/aug-cc-pVDZ (FIG. \ref{Figure5}(a)), with the deviation distribution centered at 4.74 eV with a standard deviation of 0.32 eV. After increasing the basis set to aug-cc-pVTZ, the deviation distribution centered at 2.97 eV with a standard deviation of 0.21 eV (FIG. \ref{Figure5}(b)). This proves that a large basis set is necessary to calculate core-excited states. When adopting the CVS-MRCI scheme in conjunction with the aug-cc-pVDZ basis set, the deviation distribution centered at 2.92 eV with a standard deviation of 0.20 eV  indicated that the significance of the dynamic correlation in the calculation of core-excited states. When combined with the aug-cc-pVTZ basis set, the deviation distribution centered at -0.09 eV with a standard deviation of 0.14 eV, as illustrated in FIG. \ref{Figure5}(f), demonstrating that the CVS-MRCI scheme combined with a large basis set can produce calculated results that are consistent with the experimental values. In previous reports, relativistic effect corrections to the calculation results often yielded better results. Considering the relativistic correction, the centers of the deviation distributions between the calculated results of CVS-MCSCF/aug-cc-pVDZ-DK (FIG. \ref{Figure5}(c)), CVS-MCSCF/aug-cc-pVTZ-DK (FIG. \ref{Figure5}(d)) and the experimental values are situated at 5.02 and 3.32 eV, respectively. While the centers of the deviation distributions between the results of CVS-MRCI/aug-cc-pVDZ-DK (FIG. \ref{Figure5}(g)), CVS-MRCI/aug-cc-pVTZ-DK (FIG. \ref{Figure5}(h)) and the experimental values are 3.19 eV and 0.21 eV, respectively. 
	
\begin{figure}[htbp]
		\includegraphics{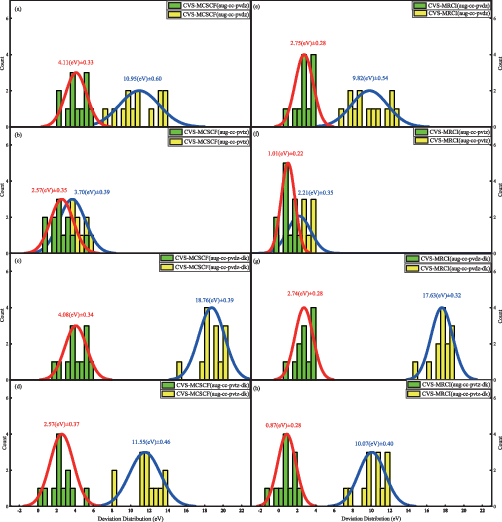}%
		\caption{\label{Figure7} The deviation distribution in the computed excitation energies for the second-row elements (Si, P, S, and Cl). The blue and red curves describe the normal distribution of the excitation energy deviation of the $1\textit{s}$ and $2\textit{p}$ orbitals, respectively.}
	\end{figure}

	The calculated core-excited energies for excitations from $1\textit{s}$ and $2\textit{p}$ orbitals localized on the second-row elements (Si, P, S, and Cl) are listed in Table \ref{SI:table2} \cite{SI}. The deviation between the calculated and experimental values, as well as the deviation distributions are illustrated in FIG. \ref{Figure6} and FIG. \ref{Figure7}, respectively. The CVS-MCSCF results demonstrate an overestimation of the calculated excitation energies that worsens as relativistic effects are taken into account, as shown in FIG. \ref{Figure7}(a-d).  Compared to CVS-MCSCF, CVS-MRCI/aug-cc-pVTZ yields more accurate results, as shown in FIG. \ref{Figure7}(e-h), which are comparable to those for the first-row elements. However, the deviation between the calculated and experimental values rises for both methods compared to the first-row elements. Although the error is small in percentage terms, accurate predictions of the absolute excitation energy are helpful in interpreting experimental spectra. For the second-row elements, corrections for the relativistic effect become more important. As shown in FIG. \ref{Figure6}, it is clear that the correct ratio of the relativistic effect to the vertical excitation energies of the  $1\textit{s}$ electrons is more than that of the  $2\textit{p}$ electrons. As an illustration, the deviation is doubled when the relativistic effect correction is taken into account to the CVS-MCSCF calculation of the vertical excitation energy of the $1\textit{s}$ electron of the Cl element.

	The normal distribution of the excitation energy deviation of the $1\textit{s}$ and $2\textit{p}$ electrons in the second-row elements (Si, P, S and Cl) are represented by the blue and red curves in FIG. \ref{Figure7}. According to FIG. \ref{Figure7}(a, b), CVS-MCSCF/aug-cc-pVDZ performs marginally worse in describing the excited states of $2\textit{p}$ electrons, with the deviation distribution centered at 4.11 eV, which is 1.5 eV higher than that with aug-cc-pVTZ basis set. Worse, the center of the deviation distribution between the $1\textit{s}$ excited electrons states calculated by CVS-MCSCF/aug-cc pVDZ and the experimental value is 10.95 eV. In comparison, the center of deviation between the results calculated by CVS-MCSCF/aug-cc-pVTZ and the experimental value is at 3.70 eV, which is only 1/3 of the former. As shown in Figure 5(b, f), CVS-MRCI/aug-cc-pVTZ performs much better than CVS-MCSCF/aug-cc-pVTZ at describing the core-excited states of the second-row elements, with deviation distributions that are almost half as large and centering at 2.21 and 1.01 eV for $1\textit{s}$ and $2\textit{p}$ electrons, respectively. Table S2 displays the precise values of the vertical excitation energy of the inner electrons as calculated by CVS-MRCI/aug-cc-pVTZ. In accordance with the above, it can be concluded that the vertical excitation energy of the inner electrons of the elements in the second row can be reproduced more accurately with CVS-MRCI/aug-cc-pVTZ.
	
	\subsection{\label{sec:level2}Multi-configuration and multi-electron excitation characteristic}
	
	Although the core orbitals are localized around the corresponding atoms, the frontier orbitals are diffuse and, in some cases, almost degenerate. Strong static correlation is present in these core-excited states. MR approaches are necessary to better describe a system with a strong static correlation. 
	
	Our results show that the multi-electron excitation characteristic is widespread in the core-excited states. For example, in the X-ray absorption spectrum of an ozone molecule, the excitation energy of an electron in the $1\textit{s}$ orbital of the terminal O atom ($O_t$ for short) is 529.1 eV. According to our CVS-MRCI calculations, it is a two-electron excited state, while electrons excited from $O_t$ ($1\textit{s}$) and 9a' to 3a” respectively. This multi-electron excitation property is more obvious in the core-excited state of the $O_2$ molecule. The experimentally measured electronic excitation energy of the $1\textit{s}$ orbital of the O atom in $O_2$ is 530.8 eV, which corresponds to the electronic excitation from the $1\textit{s}$ orbital to the $2\textit{p}$ orbitals. Our calculation results confirm this fact. However, the experimental interpretation states that the absorption peak with an excitation energy of 538.9 eV corresponds to an electron from the O $1\textit{s}$ orbital to the $3\textit{s}_{\sigma g}$ orbital. However, our calculations show that this absorption peak corresponds to electrons from the $1\textit{s}$ and 2$b_1$ orbitals to the 2$b_2$ and 6$a_1$ orbitals, respectively. This property of multi-electron excitation is widespread in the core-excited states of these small molecules, which means that the development and application of the multi-configuration approach to explain the core electron excited states is crucial.
	
	\subsection{\label{sec:level2}Core Electron Vertical Ionization Potentials (CEVIPs)}
	
	\begin{figure}[htbp]
		\includegraphics{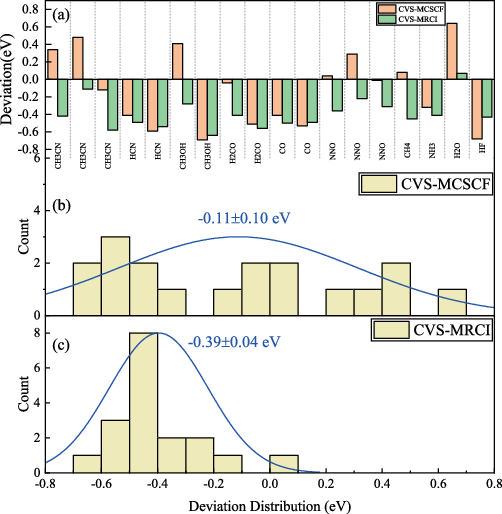}%
		\caption{\label{Figure8} Deviation of the calculated core electron binding energies from the experimental values for the first-row elements (C, N, O, and F). And the deviation distribution of core electron binding energies was calculated using the CVS-MCSCF and CVS-MRCI methods for the first-row elements, respectively.}
	\end{figure}
	Theoretical simulations of CEVIPs are necessary for the interpretation of experimental X-ray photoelectron spectra, which can also reveal information about chemical interactions and chemical bonds in gas phases. Previous X-ray and synchrotron observations are included in the experimental data, as listed in Table S3. CEVIPs can be calculated as the energy difference between the ground state of the neutral molecule and the core-ionized cation. Additionally, the ground state structures of the neutral molecule are optimized by the MCSCF/aug-cc-pVTZ scheme. Table S3 contains the calculated CEVIPs for the first-row elements (C, N, O, and F) in small molecules. The deviation between the calculated and experimental CEVIPs as well as the deviation distribution is illustrated in FIG. \ref{Figure8}. The CEVIPs calculated using CVS-MCSCF/aug-cc-pVTZ have a wide deviation distribution when compared to CVS-MRCI/aug-cc-pVTZ, as shown in FIG. \ref{Figure8}(b). The core ionization energy deviation center shifts from -0.11 eV with a standard deviation of 0.10 eV to -0.39 eV with a standard deviation of 0.04 eV. A lower standard deviation indicates the stability of the CVS-MRCI results. From FIG. \ref{Figure8}(a), it is clear that the CEVIPs calculated using the CVS-MRCI/aug-cc-pVTZ scheme are systematically underestimated. We believe this is caused by the varied active spaces that were selected when calculating the ground and core-excited states as well as the various frontier orbital optimization levels when CVS-MCSCF is used as the core electron ionization state.
	
	\section{\label{sec:level1}CONCLUSIONS}
	
	A practical multi-reference configuration interaction approach for core-level excitation spectra and ionization potentials is proposed by introducing the core-valence separation approximation in GUGA-based MRCISD. The CVS approximation with GUGA-based methods can be implemented by flexible truncation of DRT. To prevent variational collapse during the diagonalization of the Davidson iteration, all non-core excitation configurations must be eliminated from both the reference and excitation configuration spaces. Our results show that the CVS-MRCI/aug-cc-pVTZ is capable of reproducing core-excited energies that are consistent with the experiments. The deviation center of the calculated vertical excitation energy of $1\textit{s}$ electron for the first-row elements is only -0.09 eV with a standard deviation of 0.14 eV. The centers of deviation for calculating the vertical excitation energy of the $1\textit{s}$ electron and the $2\textit{p}$ electron for the second-row elements are 2.21 eV with a standard deviation of 0.35 eV and 1.01 eV with a standard deviation of 0.22 eV, respectively. The results confirm the fact that the dynamic correlation between electrons makes an undeniable contribution in core-excited states, compared with the calculation results of CVS-MCSCF.

	\begin{acknowledgments}
		This work was supported by the National Natural Science Foundation of China under Grants (No. 11904287, 11847041, 22273071), and the Double First-class University Construction Project of Northwest University.
	\end{acknowledgments}

	\section{DEDICATION}
	
	In memory of Prof. Zhenyi Wen, who passed away on the May 2, 2022. He was an outstanding scientist with significant contributions to many areas of quantum chemistry, particularly in the development of the GUGA-based multi-reference configuration interaction (MRCI) method.

	\nocite{*}
	
	\bibliography{Manuscript}
	
\end{document}


\preprint{APS/123-QED}

\title{GUGA-based MRCI approach with Core-Valence Separation Approximation (CVS) for the calculation of the Core-Excited States of molecules}

\author{Qi Song~\orcidlink{0000-0003-4851-3940}}
\email{songq@nwu.edu.cn}%
\author{ Baoyuan Liu}
\author{ Junfeng Wu}
\author{ Wenli Zou~\orcidlink{0000-0002-2308-2294}}
\author{ Yubin Wang}
\author{ Bingbing Suo~\orcidlink{0000-0002-2308-2294}}
\email{bsuo@nwu.edu.cn}
\affiliation{1.Institute of Modern Physics, Northwest University, Xi’an, Shaanxi 710069, China}
\author{ Yibo Lei~\orcidlink{0000-0002-0747-2428}}
\email{leiyb@nwu.edu.cn}
\affiliation{2.College of Chemistry and Materials Science, Northwest University, Xi’an, Shaanxi 710069, People’s Republic of China}

\date{\today}

\maketitle
\renewcommand\thetable{S\arabic{table}}
\renewcommand\thefigure{S\arabic{figure}}
\setcounter{table}{0}
\setcounter{figure}{0}

\section{\label{sec:level1}SUPPLEMENTAL MATERIAL}

\clearpage
\begin{table*}[htbp]
	\caption{\label{SI:table1}Computed vertical excited energies (in eV) for the first-row atoms (C, N, O, and F) with multi-reference CVS-MCSCF and CVS-ic-MRCI methods.}
	\begin{ruledtabular}
		\begin{tabular}{lp{1cm}p{1cm}p{1cm}p{1cm}p{1cm}p{1cm}p{1cm}p{1cm}p{1cm}}
			&&\multicolumn{4}{c}{CVS-MCSCF} & \multicolumn{4}{c}{CVS-ic-MRCI} \\
			Excitation& Exp. & basis set\footnotemark[1]$& basis set\footnotemark[2]$& basis set\footnotemark[3]& basis set\footnotemark[4]& basis set\footnotemark[1]$& basis set\footnotemark[2]$& basis set\footnotemark[3]& basis set\footnotemark[4]\\ \hline
			\textbf{C}{\rm O C(1s)→2b1} (0.82)&287.4&290.94&289.44&290.84&289.55&289.36&287.33&289.93&287.43\\
			\textbf{C}{\rm O C(1s)→7a1} (0.64)&292.4&297.41&295.99&297.51&296.09&294.73&292.21&294.83&292.31 \\
			\multirow{2}{*}{\textbf{C}{\rm O C(1s)→6a1}} &293.3&\multirow{2}{*}{298.70}&\multirow{2}{*}{297.19}&\multirow{2}{*}{298.81}&\multirow{2}{*}{297.29}&\multirow{2}{*}{296.08}&\multirow{2}{*}{293.46}&\multirow{2}{*}{296.18}&\multirow{2}{*}{293.56}\\
			&293.5&&&&&&&&\\
			C\textbf{O}{\rm \: O(1s)→2b1} (0.82)&534.2&538.69&536.23&537.27&536.62&537.28&534.15&537.67&534.54\\
			C\textbf{O}{\rm \: O(1s)→7a1} (0.56)&538.9&544.35&542.69&544.74&543.08&542.06&538.99&542.44&539.37\\
			C\textbf{O}{\rm \: O(1s)→6a1} (0.52)&539.9&545.41&543.73&545.81&544.12&543.20&540.10&543.59&540.48\\
			\textbf{C}O_2{\rm \:C(1s) → 3b1} (0.77)&290.8&294.38&292.95&294.48&293.06&293.40&291.04&293.51&291.14\\
			\textbf{C}O_2{\rm \:C(1s) → 8a1} (0.70)&292.7&296.55&294.85&296.65&294.94&295.12&292.72&295.21&292.81\\
			\textbf{C}O_2{\rm \:C(1s) →} &295.0&-&-&-&-&-&-&-&-\\
			\textbf{C}H_4{\rm \:C(1s) → 3a1} (0.71)&287.1&289.78&288.57&289.89&288.68&288.82&286.62&288.93&286.72\\
			\textbf{C}H_4{\rm \:C(1s) → 3b1} (0.72)&288.0&290.99&289.34&291.10&289.44&290.18&287.92&290.28&288.02\\
			\multirow{2}{*}{\textbf{C}_2{\rm H_2 \: C(1s) → 2b1} (0.71)}& \multirow{2}{*}{285.8} &290.48&288.54&290.59&288.65&288.56&286.09&288.67&286.20\\
			&&290.52&288.58&290.63&288.69&288.61&286.13&288.72&286.24\\
			\multirow{2}{*}{\textbf{C}_2{\rm H_2 \: C(1s) → 8a1} (0.65)}& \multirow{2}{*}{287.7} &290.69&289.51&290.8&289.61&290.11&287.72&290.22&287.82\\
			&&290.76&289.58&290.87&289.65&290.19&287.8&290.3&287.90\\
			\multirow{2}{*}{\textbf{C}_2{\rm H_2 \: C(1s) → 6a1} (0.66)}& 288.7&291.17&290&291.28&290.11&290.4&288.04&290.51&288.14\\
			&288.8&291.24&290.07&291.35&290.18&290.48&288.12&290.59&288.22\\
			H_2\textbf{C}{\rm O \: C(1s) → 2b1} (0.76)&286.0&290.10 &288.68 &290.21 &288.78 &288.52 &286.27 &288.63 &286.38 \\
			H_2\textbf{C}{\rm O \: C(1s) → 6a1} (0.64)&290.2&294.37 &293.23 &294.48 &293.34 &291.76 &289.54 &291.86 &289.64 \\
			H_2\textbf{C}{\rm O \: C(1s) → 7a1} (0.62)&291.3&295.84 &294.70 &295.95 &294.81 &293.11 &290.89 &293.22 &290.99 \\
			H_2{\rm C\textbf{O} \: O(1s) → 2b1} (0.51)&530.8&535.65 &533.24 &536.05 &533.63 &533.74 &530.40 &534.14 &530.79 \\
			H_2{\rm C\textbf{O} \: O(1s) → 6a1} (0.49)&535.4&540.38 &538.42 &540.78 &538.81 &538.23 &534.96 &538.62 &535.34 \\
			H_2{\rm C\textbf{O} \: O(1s) → 3b2} (0.62)&536.3&540.61 &538.07 &541.00 &538.45 &539.20 &535.85 &539.59 &536.24 \\
			Double Exited (0.51)&538.0&543.35 &540.96 &543.75 &541.35 &540.83 &537.50 &541.22 &537.89 \\
			\multirow{2}{*}{\textbf{N}_2{\rm \: N(1s) → 2b2} (0.67)}& \multirow{2}{*}{401.0} &405.90&404.25&406.11&404.46&403.64&400.74&403.86&400.95\\
			& &406.15&404.68&406.37&404.9&404.01&401.12&404.23&401.33\\
			\multirow{2}{*}{\textbf{N}_2{\rm \: N(1s) → 7a1} (0.62)}& \multirow{2}{*}{406.2} &411.57&410.73&411.78&410.14&409.42&406.46&409.62&406.55\\
			&&414.34&410.86&414.55&413.33&410.06&406.81&410.26&407.33\\
			\multirow{2}{*}{\textbf{N}_2{\rm \: N(1s) → 6a1} (0.64)}& 407.1&415.80&411.39&416.03&411.62&412.39&407.2&412.61&407.42\\
			&407.3&416.68&411.62&416.91&411.85&413.16&407.55&413.39&407.78\\
			\textbf{N}NO N(1s) → 3b1 (0.70)&401.1&406.04&403.95&406.26&404.53&404.38&401.57&404.60&401.88\\
			\multirow{2}{*}{\textbf{N}NO N(1s) → 8a1 (0.50)}&404.0&408.21&406.59&408.42&407.10&406.87&404.18&407.08&404.38\\
			&405.9&-&-&-&-&-&-&-&-\\
			N\textbf{N}O N(1s) → 2b1 (0.69)&404.8&409.77&407.83&409.99&408.04&407.93&405.31&408.14&405.52 \\
			N\textbf{N}O N(1s) → 8a1 (0.57)&407.5&413.96&411.99&414.17&412.20&412.28&409.59&412.49&409.80\\
			N\textbf{N}O N(1s) → 8a1 (0.57)&407.8&414.24&412.21&414.45&412.42&412.59&409.81&412.80&410.93\\
			\multirow{2}{*}{\textbf{O}_2{\rm \: O(1s) → 2b1} (0.70)}& \multirow{2}{*}{530.8} &534.81&533.81&535.21&534.21&533.27&530.18&533.67&530.56\\
			&&535.20&534.21&535.59&534.60&533.74&530.66&534.13&531.05\\
			\textbf{O}_2{\rm \: O(1s) → 6a1} (0.59)&538.9&542.77&541.97&543.16&542.35&541.47&538.64&541.85&539.01\\
			\multirow{2}{*}{Double Excited (0.70)}& \multirow{2}{*}{539.9} &544.61&543.73&544.99&544.10&542.49&539.46&542.86&539.83\\
			&&544.95&544.07&545.33&544.44&542.79&539.84&543.20&540.24\\
			\textbf{O}_3{\rm \: Oc(1s)-3a”} (0.79)&535.4&538.28&536.93&538.67&537.31&537.24&534.35&537.63&534.73\\
			\textbf{O}_3{\rm \: Oc(1s)-11a’} (0.51)&542.3&543.61&542.38&543.98&542.74&543.34&540.56&543.71&540.92\\
			\textbf{O}_3{\rm \: Oc(1s)-12a”} (0.53)&544.0&546.19&544.93&546.57&545.30&545.62&542.82&545.99&543.19\\
			\textbf{O}_3{\rm \: Ot(1s)- 3a”} (0.67)&529.1&532.54&531.11&532.94&531.50&531.96&528.92&532.36&529.31\\
			Double Excited&-&535.32&533.85&535.72&534.24&534.57&531.47&534.94&531.86\\
			\textbf{O}_3{\rm \: Ot(1s)- 11a’} (0.57)&536.3&540.92&539.46&541.31&539.85&539.53&536.69&539.91&537.05\\
			
		\end{tabular}
	\end{ruledtabular}
\footnotetext[1]{aug-cc-pVDZ}
\footnotetext[2]{aug-cc-pVTZ}
\footnotetext[3]{aug-cc-pVDZ-DK}
\footnotetext[4]{aug-cc-pVTZ-DK}
\end{table*}

\begin{table*}[htbp]
	\caption{\label{SI:table2}Computed vertical excited energies (in eV) for the second-row atoms (Si, P, S, and Cl) with multi-reference CVS-MCSCF and CVS-ic-MRCI methods.}
	\begin{ruledtabular}
		\begin{tabular}{lp{1cm}p{1cm}p{1cm}p{1cm}p{1cm}p{1cm}p{1cm}p{1cm}p{1cm}}
			&&\multicolumn{4}{c}{CVS-MCSCF} & \multicolumn{4}{c}{CVS-ic-MRCI} \\
			Excitation& Exp. & basis set\footnotemark[1]$& basis set\footnotemark[2]$& basis set\footnotemark[3]& basis set\footnotemark[4]& basis set\footnotemark[1]$& basis set\footnotemark[2]$& basis set\footnotemark[3]& basis set\footnotemark[4]\\ \hline
			\textbf{Si}H_4 \:{\rm Si(1s)→6a1} (0.74)&1842.5&1856.30&1846.55&1860.70&1850.89&1855.27&1845.51&1859.67&1849.85 \\
			\textbf{Si}H_4 \:{\rm Si(2s)→}&-&158.19&156.81&158.88&157.49&157.14&155.75&157.82&156.43 \\
			\textbf{Si}H_4 \:{\rm Si(2p)→6a1} (0.77)&102.8&106.25&104.60&106.20&104.53&105.08&102.80&105.02&103.37 \\
			H_2\textbf{S} \:{\rm S(1s) → 3b2} (0.83)&2473.1&2483.60&2476.83&2491.42&2484.61&2483.13&2475.90&2490.95&2483.67 \\
			H_2\textbf{S} \:{\rm S(1s) → 6a1} (0.68)&2476.3&2483.63&2476.95&2491.42&2484.70&2483.07&2476.05&2490.87&2483.80 \\
			H_2\textbf{S} \:{\rm S(2s) →} &-&232.04&230.80&233.37&232.13&231.40&229.78&232.89&231.11 \\
			H_2\textbf{S} \:{\rm S(2s) →} &-&232.28&231.20&233.59&232.50&231.58&230.21&232.74&231.51 \\
			H_2\textbf{S} \:{\rm S(2p) → 3b2}(0.85)&164.4&168.44&166.85&168.43&166.83&167.78&165.82&167.77&165.81 \\
			H_2\textbf{S} \:{\rm S(2p) → 6a1}(0.83)&166.5&168.86&167.43&168.83&167.39&168.08&166.38&168.05&166.34 \\
			\textbf{S}O_2 \:{\rm S(1s) → 4a"} (0.83)&2473.8&2484.47&2478.04&2492.28&2485.80&2483.42&2476.54&2491.23&2484.31 \\
			\textbf{S}O_2 \:{\rm S(1s) → 13a"} (0.73)&2478.4&2488.18&2481.93&2495.98&2489.65&2486.79&2480.03&2494.49&2487.81 \\
			Double Excited&2478.9&2491.34&2483.63&2499.14&2492.47&2489.82&2481.62&2497.63&2490.58 \\
			\textbf{S}O_2 \:{\rm S(2p) → 14a"} (0.65)&171.3&173.30&171.80&173.15&171.77&171.87&170.01&171.85&169.99 \\
			\textbf{S}O_2 \:{\rm S(2p) → 4a"} (0.78)&164.4&169.45&167.83&169.43&167.81&168.21&166.21&168.19&166.19 \\
			\textbf{P}H_3 \:{\rm P(1s) →8a'} (0.81)&\multirow{3}{*}{2145.8}&2158.94&2150.95&2164.86&2156.82&2157.75&2149.5&2163.66&2155.38 \\
			\textbf{P}H_3 \:{\rm P(1s) →9a'} (0.74)&&2159.18&2151.06&2165.11&2156.94&2157.84&2149.43&2163.77&2155.31 \\
			\textbf{P}H_3 \:{\rm P(1s) →3a"} (0.84)&&2159.5&2151.43&2165.43&2157.31&2157.89&2149.57&2163.82&2155.43 \\
			\textbf{P}H_3 \:{\rm P(2p) → 3a"}(0.83)&\multirow{3}{*}{132.3}&136.14&134.4&136.11&134.36&134.82&132.83&134.79&132.79 \\
			\textbf{P}H_3 \:{\rm P(2p) → 8a'}(0.83)&&136.15&134.69&136.11&134.64&134.82&132.85&134.8&132.82 \\
			\textbf{P}H_3 \:{\rm P(2p) → 9a'} (0.49)&&136.18&133.14&136.14&134.66&134.94&133.14&134.89&132.98 \\
			H\textbf{Cl} \:{\rm Cl(1s) → 6a1} (0.66)&2823.9&2834.14&2827.63&2844.24&2837.68&2832.42&2825.49&2842.52&2835.54 \\
			H\textbf{Cl} \:{\rm Cl(1s) → 7a1} (0.66)&\multirow{2}{*}{2827.8}&2837.12&2830.31&2847.24&2840.38&2835.61&2828.44&2845.73&2838.51 \\
			H\textbf{Cl} \:{\rm Cl(1s) → 3b1} (0.88)&&2837.75&2830.87&2847.86&2840.94&2836.57&2829.38&2846.68&2839.45 \\
			H\textbf{Cl} \:{\rm Cl(2s) →}&-&275.43&273.73&277.64&275.49&273.75&272.15&275.56&273.91 \\
			\multirow{2}{*}{H\textbf{Cl} \:{\rm Cl(2s) →}}&\multirow{2}{*}{-}&278.14&276.12&280.18&277.89&276.65&274.85&278.47&276.62 \\
			&&279.04&277.56&280.81&279.33&277.71&276.01&279.49&277.77 \\
			H\textbf{Cl} \:{\rm Cl(2p) → 6a1} (0.65)&201.0&206.64&204.98&206.64&204.97&204.60&202.81&204.60&202.80 \\
			H\textbf{Cl} \:{\rm Cl(2p) → 7a1} (0.67)&\multirow{2}{*}{204.6}&209.20&207.29&209.21&207.30&207.44&205.4&207.45&205.41 \\
			H\textbf{Cl} \:{\rm Cl(2p) → 3b1} (0.88)&&209.79&208.07&209.80&208.07&208.39&206.41&208.40&206.42\\
			\textbf{Cl}_2 \:{\rm Cl(1s) → 18a'} (0.78)&2821.3&2829.36&2823.15&2839.46&2833.21&2829.24&2822.47&2839.35&2832.52 \\
			\textbf{Cl}_2 \:{\rm Cl(2s) →18a'}&-&265.64&264.58&267.36&266.29&265.30&263.73&267.03&265.44 \\
			\textbf{Cl}_2 \:{\rm Cl(2p) → 18a'} &198.7&204.13&203.71&204.13&203.70&202.21&200.77&202.21&200.76 \\
		\end{tabular}
	\end{ruledtabular}
	\footnotetext[1]{aug-cc-pVDZ}
	\footnotetext[2]{aug-cc-pVTZ}
	\footnotetext[3]{aug-cc-pVDZ-DK}
	\footnotetext[4]{aug-cc-pVTZ-DK}
\end{table*}

\begin{table}[htbp]
\caption{\label{SI:table3}
A table with numerous columns that still fits into a single column. 
Computed Ionization energies (the errors in parenthesis, eV) for C and N 1s in small molecules with multi-reference CVS-MCSCF and CVS-ic-MRCI methods. Bold type denotes the atom at which the 1\textit{s} electron is ionized}
\begin{ruledtabular}
\begin{tabular}{cccc}
Molecule (gas) & Exp.(eV) & CVS-MCSCF^{a} &CVS-ic-MRCI^{a}  \\
\hline
$\textbf{C}{\rm H_{3}CN}$&292.98&293.32 (0.34)&292.56 (-0.42)\\
CH_3{\rm \textbf{C}N}&292.45&292.93 (0.48)&292.34 (-0.11)\\
CH_3{\rm C\textbf{N}}&405.64&405.52 (-0.12)&405.06 (-0.58)\\
H\textbf{C}N&293.40&292.99 (-0.41)&292.91 (-0.49)\\
HC\textbf{N}&406.78&406.19 (-0.59)&406.24 (-0.54)\\
\textbf{C}H_3{\rm OH}&292.42&292.83 (0.41)&292.14 (-0.28)\\
CH_3{\rm \textbf{O}H}&539.11&538.42 (-0.69)&538.47 (-0.64)\\
H_2{\rm \textbf{C}O}&294.47&294.43 (-0.04)&294.06 (-0.41)\\
H_2{\rm C\textbf{O}}&539.48&538.97 (-0.51)&538.92 (-0.56)\\
\textbf{C}O&296.21&295.80 (-0.41)&295.71 (-0.50)\\
C\textbf{O}&542.55&542.02 (-0.53)&542.06 (-0.49)\\
\textbf{N}NO&408.71&408.75 (0.04)&408.35 (-0.36)\\
N\textbf{N}O&412.59&412.88 (0.29)&412.37 (-0.22)\\
NN\textbf{O}&541.42&541.41 (-0.01)&541.11 (-0.31)\\
\textbf{C}H_4&290.91&290.99 (0.08)&290.46 (-0.45)\\
\textbf{N}H_3&405.56&405.24 (-0.32)&405.15 (-0.41)\\
H_2{\rm \textbf{O}}&539.40&540.04 (0.64)&539.47 (0.07)\\
H\textbf{F}&694.23&693.55 (-0.68)&693.80 (-0.43)\\
\end{tabular}
\end{ruledtabular}
\footnotetext[a]{aug-cc-pVTZ}
\end{table}

\clearpage
\begin{figure}[htbp]
	\begin{center}
		\includegraphics[angle=90,width=0.9\textwidth]{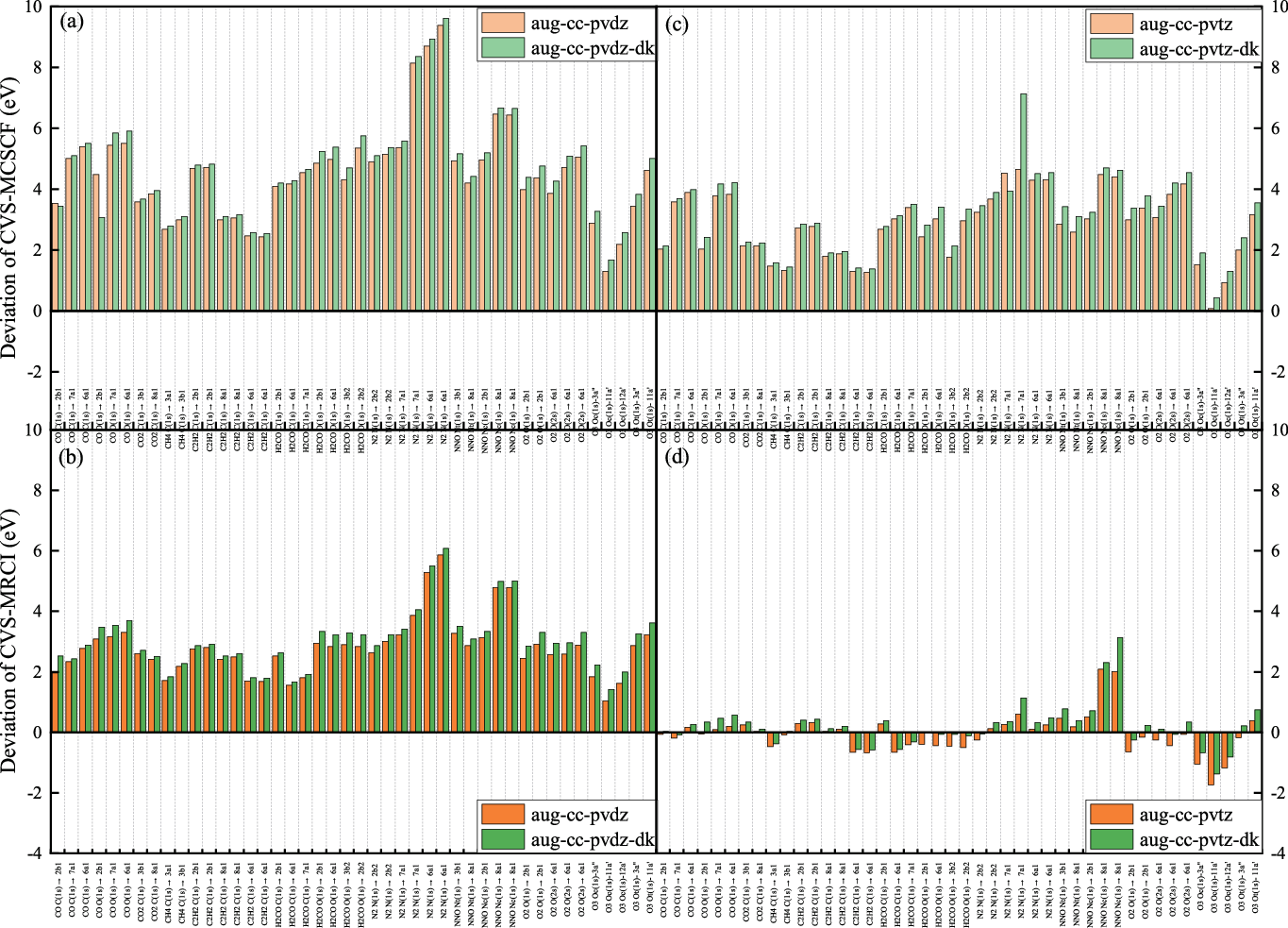}%
		\caption{Deviation of the calculated core-excited energies from the experimental values for the first-row elements (C, N, O, and F). The deviation between the experimental value and the excitation energies calculated by CVS-MCSCF with aug-cc-pVDZ (a), and aug-cc-pVTZ (b), respectively. And the deviation between the experimental value and the excitation energies calculated by CVS-MRCI with aug-cc-pVDZ (c), and aug-cc-pVTZ (d), respectively.}
		\label{Figure4}
	\end{center}
\end{figure}

\begin{figure}[htbp]
	\begin{center}
		\includegraphics[angle=90,width=0.9\textwidth]{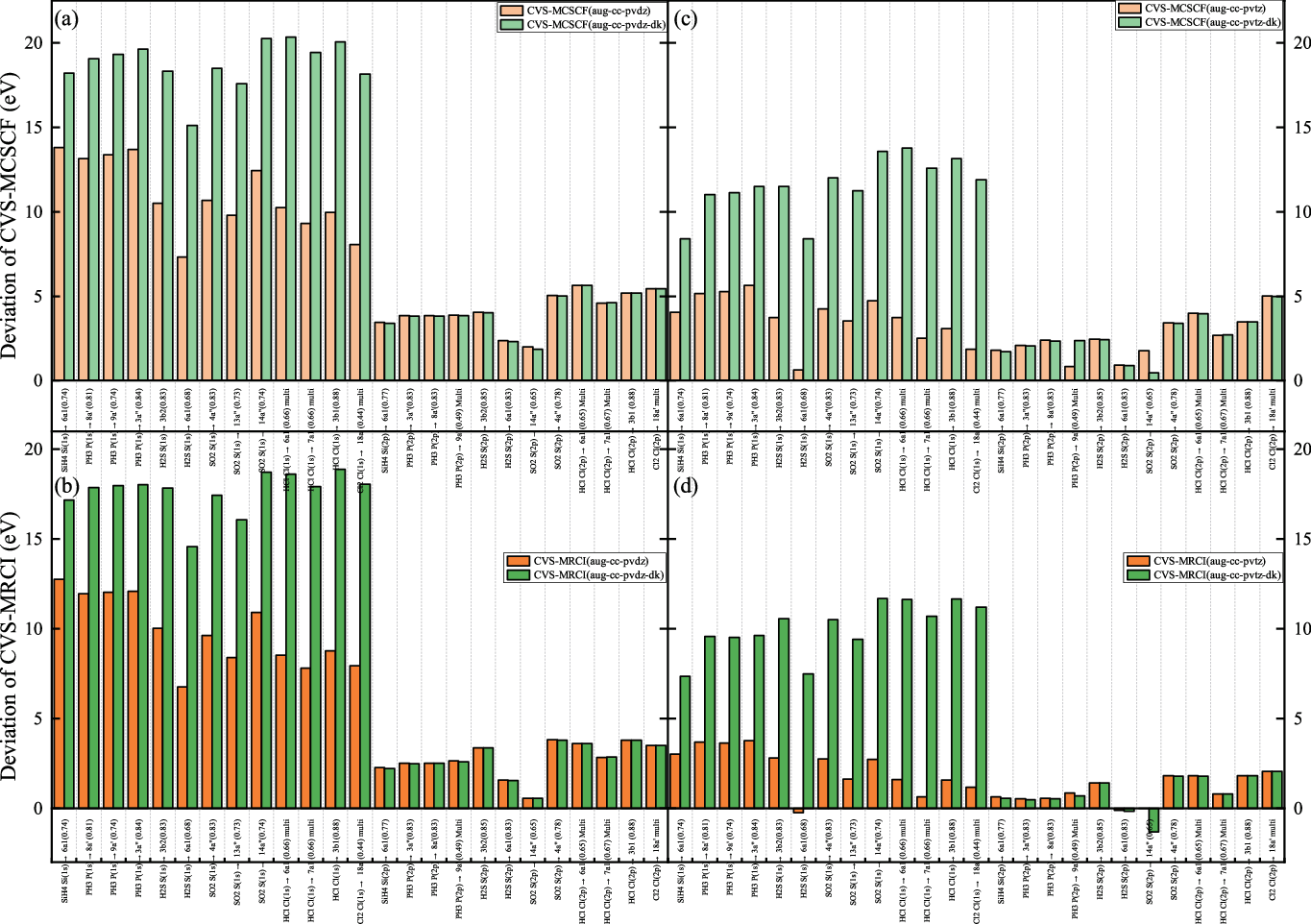}%
		\caption{Deviation of the calculated core-excited energies from the experimental values for the second-row elements (Si, P, S, and Cl).}
		\label{Figure6}
	\end{center}
\end{figure}